\documentclass[]{mn2e}
\usepackage{graphicx,mathptm,amsmath,amssymb}

\def\pd#1#2{{\upartial #1 \over \upartial #2}}

\def\sspd#1#2{{\upartial ^2 #1 \over \upartial #2^2}}
\def\tfrac#1#2{{\textstyle\frac{#1}{#2}}}
\def\vect#1{{\mathbfit{#1}}}

\def\bg#1{{\mbox{\bf $ #1 $}}}
\title
[Modal decomposition of astronomical images]
{Modal decomposition of astronomical images with application to shapelets}
\author[R.H.~Berry, M.P.~Hobson \& S.~Withington]
{R.H.~Berry, M.P.~Hobson and S.~Withington\\ 
Astrophysics Group,
Cavendish Laboratory, Madingley Road, Cambridge, CB3 0HE, UK}
\date{Accepted ---. Received ---; in original form \today}
\pagerange{\pageref{firstpage}--\pageref{lastpage}}
\pubyear{2002}

\begin{document}
\maketitle
\label{firstpage}

\begin{abstract}
The decomposition of an image into a linear combination of digitised
basis functions is an everyday task in astronomy. A general method is
presented for performing such a decomposition optimally into an
arbitrary set of digitised basis functions, which may be linearly
dependent, non-orthogonal and incomplete. It is shown that such
circumstances may result even from the digitisation of continuous
basis functions that are orthogonal and complete.  In particular,
digitised shapelet basis functions are investigated and are shown to
suffer from such difficulties. As a result the standard method of
performing shapelet analysis produces unnecessarily inaccurate
decompositions. The optimal method presented here is shown to yield
more accurate decompositions in all cases.
\end{abstract}


\section{Introduction}
\label{intro}

The linear decomposition of an astronomical dataset into a series of
basis functions is a fundamental task in many areas of astrophysics
and cosmology. Indeed, in the analysis of one-dimensional spectra,
two-dimensional images or higher-dimensional datasets, one is often
faced with the problem of determining the coefficients in a linear
expansion of the dataset in some set of (sampled) basis functions.  To
illustrate our discussion, we will focus here on the important specific
example of decomposing a two-dimensional astronomical image, although
the general approach that we advocate will be applicable to
datasets of arbitrary dimensionality.

The description of a digitised image as a linear combination of a set
of sampled basis functions is an everyday problem for astronomers.
For example, one often describes an image in terms of
a set of orthogonal Fourier modes by calculating the coefficients in
the expansion using a discrete Fourier transform. Significant
attention has also been given to representing an image as a linear
combination of discrete wavelets basis functions, both orthogonal and
non-orthogonal (see, for instance, 
Hobson et al. 1999; Sanz et al 1999a;b; Tenorio et
al. 1999). More recently, several authors have
investigated the use of Gaussian-Hermite modes (or shapelets) in
representing images of galaxies (Refregier 2003; Refregier \& Bacon 2003).  
Although such image decompostion
is commonplace, there exist a number of subtleties in the
procedure that are not widely appreciated within the astronomical
community. These include the effect of digitisation, or sampling, on
familiar notions from the analytic theory of continuous functions, in
particular the
concepts of completeness and linear independence of sampled basis
functions, and the use of non-orthogonal bases.

In this paper, we therefore discuss a general mathematical framework
for the linear decomposition of an astronomical image into a set of
arbitrary sampled functions (or modes).  These may, in general, form a
`basis' that is non-orthogonal and either under-complete,
perfectly-complete or over-complete. In the under-complete case, the
modes do not support all the degrees of freedom in an arbitrary
image. By perfectly-complete we mean that the modes supports all the
degrees of freedom without linear dependence between them, whereas for
an over-complete mode set there are additional modes relative to the
complete case and hence linear dependence in the set. In particular,
we show how to quantify those degrees of freedom in an image that can
be supported by any particular mode set, and we explicitly consider
how to obtain the optimal set of mode coefficients for representing
the image. These methods present an opportunity for the optimisation
of sampling within modal analysis, ensuring that computational
operations are reduced to a minimum.

The structure of the paper is as follows. In Section~\ref{modal}, 
we outline the modal decomposition problem and present a general
prescription for obtaining optimal results, which is based on the
singular value decomposition of the matrix of digitised mode vectors.
In Section~\ref{one-dim}, we illustrate our general approach by 
investigating the decomposition of simple one-dimensional function
into Gauss-Hermite (shapelet) modes. This investigation is pursued
further in Section~\ref{imagedecomp}, where we consider the
decomposition of images of Hubble Deep Field galaxies into
two-dimensional Gauss-Laguerre (polar shapelet) modes.
Finally, our conclusions are presented in section~\ref{conc}.

\section{The modal decomposition problem}
\label{modal}

Consider a 2-D digitised image
consisting of $N$ pixels, which we denote by the
$N$-dimensional column vector $\vect{d}$. Our goal is to represent the
image, as accurately as possible, in terms of a set of $M$
(two-dimensional) modes $\{\vect{e}_k\}$ $(k=1,2,\ldots,M)$ defined at
the same points at which $\vect{d}$ is sampled; thus each $\vect{e}_k$
is also a column vector of length $N$. It is convenient to combine the
mode vectors into the single $N\times M$ mode matrix defined by
\[
\vect{E} = [\vect{e}_1\cdots\vect{e}_k\cdots\vect{e}_M].
\]
Our objective is thus to determine a suitable 
coefficient vector $\hat{\vect{a}}$ that minimises the residual
\begin{equation}
\epsilon = |\vect{E}\vect{a}-\vect{d}|.
\label{residualdef}
\end{equation}
Note that we make no requirement for the mode vectors to be mutually
orthogonal, linearly independent,  or of unit length.

Several cases may arise in this problem. There may exist a unique
coefficient vector $\hat{\vect{a}}$ that minimises
(\ref{residualdef}); this corresponds to the quantity $\epsilon^2$
possessing a single well-defined multi-dimensional quadratic minimum
in the space of coefficients $\vect{a}$. Surfaces of constant
$\epsilon^2$ are given by the Hermitian form
$(\vect{a}-\hat{\vect{a}})^\dagger \vect{R}
(\vect{a}-\hat{\vect{a}})=\mbox{constant}$, where
$\vect{R}=\vect{E}^\dagger \vect{E}$ (which is called the Gram matrix). The
eigenvectors of $\vect{R}$ determine the principal directions of this
multi-dimensional ellipsoid, and the extent along each principal
direction is inversely proportional to the corresponding
eigenvalue. If there exist approximate degeneracies in the
coefficient-space, some of the eigenvalues become very small and so
the multi-dimensional ellipsoid is considerably elongated in these
directions; this leads to a wide range of coefficients vectors
$\vect{a}$ for which $\epsilon$ takes its minimum value to within the
numerical precision. In the limiting case of exact degeneracies, some
eigenvalues are identically zero leading to an infinite extent for the
`ellipsoid' along the corresponding principal directions. In the
subspace spanned by these directions the value of $\epsilon$ takes its
minimum value precisely. In either of the last two cases, to the
numerical precision, there exist an infinite number of possible
coefficient vectors $\hat{\vect{a}}$ that minimise
(\ref{residualdef}).  The value of $\epsilon_{\rm min}$ is also of
central importance. Clearly, if $\epsilon_{\rm min}=0$, then the modal
expansion $\vect{E}\hat{\vect{a}}$ provides an exact representation of
the original image, whereas if $\epsilon_{\rm min} > 0$ it represents
only an approximation to $\vect{d}$.

Which of the above cases occurs is dependent on the completeness and
linear dependence of the mode set ${\vect{e}_k}$, and on the
particular image $\vect{d}$ being decomposed. Clearly, if the $M < N$
then the modes are `under-complete', and so cannot represent an
arbitrary image, but only an approximation to it.  This may also occur
when $M \ge N$, however, since the modes may still not form a complete
basis as a result of linear dependence between them. Nevertheless,
even with an under-complete set, the particular image $\vect{d}$ under
analysis may lie in the subspace spanned by the modes, and so may be
represented exactly. When $M \ge N$ cases arise where the modes form a complete basis, in which any image can be represented
exactly. In this case, if $M=N$ the modes must linearly independent and
form a `perfectly-complete' basis.  If $M > N$, some linear
dependence must exist between the modes and they form an
`over-complete' basis.

In general, the degree of completeness of the chosen mode set 
may not be known at the outset. Fortunately, a straightforward
technique exists for determining the degree of completeness of the mode
set, and simultaneously determining 
an `appropriate' coefficient vector $\hat{\vect{a}}$ that minimises
(\ref{residualdef}). This is achieved by performing the singular-value
decomposition (SVD) of the mode matrix $\vect{E}$, which we now discuss.

\subsection{Singular value decomposition}
\label{SVD}

The SVD of the $N\times M$ mode matrix $\vect{E}$ (which may, in
general, contain complex-valued entries) may be written (see, for
example, Golub \& Van Loan 1992)
\begin{equation}
\vect{E} = \vect{U}\bg{\Sigma}\vect{V}^\dagger,
\label{svddef}
\end{equation}
where $\vect{U}$ is unitary matrix of dimensions $N \times N$, $\vect{V}$ is
a unitary matrix of dimensions $M \times M$, and $\bg{\Sigma}$ is a
$N\times M$ matrix (the same dimensions as $\vect{E}$) that is
diagonal in the sense that $\Sigma_{ii} = \sigma_{i}$ for $i \le p$, where
$p=\mbox{min}[M,N]$, and zero otherwise. The 
coefficients $\sigma_i$ are the {\em singular values} of the matrix
$\vect{E}$.  

As is well-known, the number $r$ (say) of non-zero singular values is
equal to the rank of $\vect{E}$, which in turn is the dimensionality
of the image subspace that is spanned by the $M$ modes
$\{\vect{e}_k\}$; clearly it must be the case that $r \le M$ and $r
\le N$.  The dimensionality of the nullspace of $\vect{E}$ (the
subspace of vectors $\vect{n}$ in the coefficient space for which
$\vect{E}\vect{n}=\vect{0}$) is given by $M-r$. It is thus a simple
matter to determine the completeness of the mode set contained in
$\vect{E}$ as follows: (i) if $r < N$ the modes are under-complete;
(ii) if $r=N$ and $M=N$ the modes are perfectly complete; and (iii) if
$r=N$ and $M > N$ the modes are over-complete.  Moreover, it is
straightforward to show that the columns of $\vect{U}$ corresponding
to non-zero singular values constitute an orthonormal basis for the
range of $\vect{E}$, and the columns of $\vect{V}$ that do not
correspond to a non-zero singular value form an orthonormal basis for
the nullspace.  In practical numerical problems, it is often the case
that none of the singular values $\sigma_i$ are identically
zero. Instead, one usually sets to zero those singular values for
which $|\sigma_i|/|\sigma_1| < \eta$, where $\eta$ is some small
factor (for example, $10^{-5}$ in single precision arithmetic) and
$\sigma_1$ is the first (and largest) singular value.

Once the SVD (\ref{svddef}) has been calculated, it is also straightforward
to obtain an `appropriate' coefficient vector $\hat{\vect{a}}$ that
minimises (\ref{residualdef}). As we outline below, in all cases one should
calculate the coefficient vector using
\begin{equation}
\hat{\vect{a}} = \vect{V}\overline{\bg{\Sigma}}^\dagger\vect{U}^\dagger\vect{d},
\label{svdsoln}
\end{equation}
where the $M\times N$ matrix denoted by
$\overline{\bg{\Sigma}}^\dagger$ is constructed by taking the
transpose of $\bg{\Sigma}$ in (\ref{svddef}) and replacing each non-zero
singular value $\sigma_i$ by $1/\sigma_i$.

It is clear that, with the above construction,
$\bg{\Sigma}\overline{\bg{\Sigma}}^\dagger$ is an $N\times N$ diagonal
matrix with diagonal entries that equal unity for those values of $j$
for which $\sigma_j \neq 0$, and zero otherwise.  Using this result it
is straightforward to show that (\ref{svdsoln}) does indeed minimise
the residual (\ref{residualdef}). Modifying slightly the argument of Press et
al. (1994), suppose we were to add to $\hat{\vect{a}}$ some arbitrary
vector $\vect{a}'=\vect{a}_1' +\vect{a}_2'$, where $\vect{a}_2'$ is
the part of the vector that lies in the nullspace of $\vect{E}$ (if
one exists) and $\vect{a}_1'$ is the part that lies in the complement
to the nullspace.  This would result in the addition of the vector
$\vect{d}'=\vect{E}\vect{a}_1'$ to $\vect{E}\hat{\vect{a}}-\vect{d}$.
We would then have
\begin{eqnarray}
|\vect{E}\hat{\vect{a}}-\vect{d}+\vect{d}'|  
& = & |(\vect{U}\bg{\Sigma}\overline{\bg{\Sigma}}^\dagger
\vect{U}^\dagger-\vect{I})
\vect{d}+\vect{d}'| \nonumber \\
& = & |\vect{U} [(\bg{\Sigma}\overline{\bg{\Sigma}}^\dagger-\vect{I})
\vect{U}^\dagger\vect{d}+\vect{U}^\dagger]\vect{d}'| \nonumber \\
& = & |(\bg{\Sigma}\overline{\bg{\Sigma}}-\vect{I})
\vect{U}^\dagger\vect{d}+\vect{U}^\dagger\vect{d}'|, \label{xminres}
\end{eqnarray}
where in the last line we have made use of the fact that the length of
a vector is left unchanged under the action of the unitary matrix
$\vect{U}$. Now, the $j$th component of the vector
$(\bg{\Sigma}\overline{\bg{\Sigma}}-\vect{I})
\vect{U}^\dagger\vect{d}$ will only be non-zero when
$\sigma_j=0$. However, the $j$th element of the vector
$\vect{U}^\dagger\vect{d}'$ is non-zero only if $\sigma_j \neq 0$,
since $\vect{d}'$ lies in the range of $\vect{E}$. Thus, as these two
terms only contribute to (\ref{xminres}) for two disjoint sets of
$j$-values, its minimum value, as $\vect{a}'$ is varied, occurs when
$\vect{d}'=\vect{0}$; this requires $\vect{a}_1' =\vect{0}$.

If the image $\vect{d}$ lies in the subspace spanned by the $M$ modes
$\{\vect{e}_k\}$, then the minimum value of the residual (\ref{residualdef}) is
zero, and the image is represented exactly by
$\vect{E}\hat{\vect{a}}$. Clearly, in order to represent an arbitrary
image we require $r=N$ (and so $M \ge N$). If $\vect{d}$ does not lie
in the range of $\vect{E}$, then $\vect{E}\hat{\vect{a}}$ yields only
an approximate representation of the image (for this to occur, the
modes must be under-complete; this will always be true if $M < N$, but
may also occur when $M \ge N$).

In either case, we see from the argument above that, if $\vect{E}$
does not possess a null space, the coefficient vector (\ref{svdsoln}) is
unique in minimising the residual (\ref{residualdef}). The condition for this to
occur is that $r=M$, indicating that the $M$ modes are linearly
independent (and hence requiring $M \le N$).  If $\vect{E}$ does
possess a nullspace, however, any vector $\vect{a}_2'$ in this
nullspace can be added to (\ref{svdsoln}), without changing the value of the
residual. Thus there are an infinite number of coefficient vectors
that minimise (\ref{residualdef}). The condition for this to occur is that $r <
M$, indicating linear dependence between the modes (which may occur
for $M \le N$ and $M > N$).

In the case where ${\bf E}$ possesses a nullspace, from the infinite
number of coefficient vectors that minimise the residual (\ref{residualdef}), the
vector $\hat{\vect{a}}$ in (\ref{svdsoln}) is that which contains no
contribution from the nullspace.  An equivalent statement is that
(\ref{svdsoln}) is the vector of shortest length that minimises
(\ref{residualdef}). Consider again adding some arbitrary vector $\vect{a}'$ to
(\ref{svdsoln}). The length of the resulting vector is
\begin{eqnarray*}
|\hat{\vect{a}}+\vect{a}'| & = & 
|\vect{V}\overline{\bg{\Sigma}}^\dagger\vect{U}^\dagger\vect{d}+\vect{a}'|\\
& = & |\vect{V}(\overline{\bg{\Sigma}}^\dagger\vect{U}^\dagger\vect{d}
+\vect{V}^\dagger\vect{a}')| \\
& = & |\overline{\bg{\Sigma}}^\dagger\vect{U}^\dagger\vect{d}
+\vect{V}^\dagger\vect{a}'|, 
\end{eqnarray*}
where in the last line we again make use of the fact that the
length of a vector is left unchanged under the action of a unitary
matrix. The $j$th component
of the vector $\overline{\bg{\Sigma}}\vect{U}^\dagger\vect{d}$ 
will only be non-zero when $\sigma_j \neq 0$, whereas
the $j$th element of the vector $\vect{V}^\dagger\vect{a}'$ is non-zero
only if $\sigma_j = 0$. Thus the minimum length vector 
has $\vect{a}'=\vect{0}$.

\subsection{Dual modes}
\label{dualmodes}

Although the appropriate coefficient vector may always be
obtained straightforwardly
from (\ref{svdsoln}), it is conceptually appealing to consider coefficient
$\hat{a}_k$ as the scalar product of the image vector $\vect{d}$
with some new mode vector $\tilde{\vect{e}}_k$ that is dual to the 
original mode $\vect{e}_k$. In other words, it is often convenient to
rewrite (\ref{svdsoln}) as
\begin{equation}
\hat{\vect{a}} = \tilde{\vect{E}}^\dagger\vect{d},
\label{dualssoln}
\end{equation}
where $\tilde{\vect{E}}$ is the $N\times M$ 
dual mode matrix, which contains the
$M$ dual mode vectors as its columns, i.e.
\[
\tilde{\vect{E}} = [\tilde{\vect{e}}_1\cdots\tilde{\vect{e}}_k\cdots
\tilde{\vect{e}}_M].
\]
From (\ref{svdsoln}), we see immediately that the dual mode matrix is given
in terms of the SVD of the original mode matrix by
\[
\tilde{\vect{E}} = \vect{U}\overline{\bg{\Sigma}}\vect{V}^\dagger.
\]
Thus, once a mode set has been defined, the dual mode set can be
calculated directly by performing a SVD, without reference to any
image. The coefficients for any particular image are then quickly
obtained using (\ref{dualssoln}). 

It should be noted that, in general, 
\[
\vect{E}\tilde{\vect{E}}^\dagger \neq \vect{I}_N
\neq \tilde{\vect{E}}\vect{E}^\dagger, 
\]
where we now start to place subscripts on identity
matrices to emphasise their dimensionality. 
It is only in the case where the original mode set is
complete that 
$\bg{\Sigma}\overline{\bg{\Sigma}}^\dagger = \vect{I}_N 
=\overline{\bg{\Sigma}}\bg{\Sigma}^\dagger$, from which it
is a simple matter to verify that 
$\vect{E}\tilde{\vect{E}}^\dagger 
= \vect{I}_N = \tilde{\vect{E}}\vect{E}^\dagger.$
This, of course, corresponds to the case in which an arbitrary image $\vect{d}$
can be represented exactly, since then
\[
\vect{E}\hat{\vect{a}} = \vect{E}\tilde{\vect{E}}^\dagger \vect{d} =
\vect{d}.
\]
Otherwise only an approximation to the original image is possible.

It is also worth pointing out that, in general, 
\[
\vect{E}^\dagger\tilde{\vect{E}} \neq \vect{I}_M \neq
\tilde{\vect{E}}^\dagger\vect{E}.
\]
In other words, the dual modes do {\em not} necessarily obey the
standard orthogonality condition $\vect{e}_k^\dagger\tilde{\vect{e}}_l
= \delta_{kl} = \tilde{\vect{e}}_k^\dagger\vect{e}_l$ with the
original mode set.  This orthogonality condition is only satisfied if
the original mode set is linearly-independent (for which $r=M$ and
hence $M\le N$). In this case,
$\bg{\Sigma}^\dagger\overline{\bg{\Sigma}}= \vect{I}_M
=\overline{\bg{\Sigma}}^\dagger\bg{\Sigma}$, from which one quickly
finds that $\vect{E}^\dagger\tilde{\vect{E}} = \vect{I}_M =
\tilde{\vect{E}}^\dagger\vect{E}$, which recovers the standard result
that the dual modes form the reciprocal basis of the original mode
set.  This corresponds to the case where (\ref{dualssoln}) is unique
in minimising the residual (\ref{residualdef}).  If, in addition, we
have $M=N$ and so the basis perfectly complete, $\vect{E}$ is
invertible and it follows that $\tilde{\vect{E}}^\dagger =
\vect{E}^{-1}$. We reiterate, however, that all possible cases are
automatically accommodated using SVD.

\subsection{Orthogonal mode sets}
\label{orthogonalmodes}

So far, we have imposed no restrictions on the orthogonality,
normalisation or the number of members of our original mode set
$\{\vect{e}_k\}$. It is worth investigating, however, the
simplifications that occur in the case where the modes are 
mutually orthogonal and each mode is unique (so $M \le N$). Hence, we have
$\vect{E}^\dagger \vect{E} = \vect{I}_M$.

In this case, the mode set is automatically linearly-independent and
so the coefficient vector (\ref{svdsoln}) or (\ref{dualssoln}) is unique in minimising
the residual. Moreover, the dual modes and original modes obey the 
standard orthogonality condition 
$\vect{E}^\dagger\tilde{\vect{E}} = \vect{I}_M =
\tilde{\vect{E}}^\dagger\vect{E}$. In the case of
orthogonal modes, it is clear that, in addition, this condition is
satisfied if the dual modes are given by
\[
\tilde{\vect{e}}_k = \frac{\vect{e}_k}{|\vect{e}_k|^2}.
\]
Thus, for orthogonal modes (even with $M < N$), we see that each dual
is simply proportional to the original mode.  In the event that the
original modes are normalised to unit length, this proportionality
becomes an equality, and the dual modes are identical to the original
modes, so that $\tilde{\vect{E}} = \vect{E}$ and hence
\begin{equation}
\hat{\vect{a}} = \vect{E}^\dagger\vect{d}.
\label{standardsoln}
\end{equation}
This last result is, of course, the reason for the common practice, in
the case of orthonormal modes, of calculating the $k$th mode
coefficient $\hat{a}_k$ by simply evaluating the scalar product of the image
$\vect{d}$ with the $k$th mode $\vect{e}_k$. 
Of course, when $M=N$, a set of orthogonal modes also 
forms a perfectly complete basis.

\subsection{Practical considerations}
\label{practical}

Throughout our discussion, we have adopted the practical approach of
working with finite dimensional vectors, rather than continous
functions.  Thus, both the image $\vect{d}$ and the modes $\vect{e}_k$
are considered simply as $N$-dimensional vectors.  This allows the
direct application of our approach to a wide variety problems in
the decomposition of digital astronomical images. 
The image pixel values $d_i$ correspond to
samples of the underlying continuous distribution $d(\vect{x})$ 
at a particular set of points $\{\vect{x}_i\}$. In practice, it is
most often the case that sample points are regularly spaced, but
this restriction is not necessary. The entire approach
can just as easily be applied to the case where the pixels values
correspond to samples that are not regularly spaced. In an analogous
manner, the elements of each mode vector $\vect{e}_k$ are
the sample values from some underlying continuous mode 
function $f_k(\vect{x})$ at the same set of points $\{\vect{x}_i\}$.

It is clear that the positions of the
sample points will have a
profound effect on whether the linear-independence, orthogonality,
or other properties of the original set of continuous mode
functions $\{f_k(\vect{x})\}$ are inherited by the mode vectors
$\{\vect{e}_k\}$. Consider, for example, a set of continuous mode 
functions $\{f_k(\vect{x})\}$
that are orthonormal over some continuous domain ${\cal D}$, so that
\[
\int_{\cal D} 
f^{\ast}_k(\vect{x}) f_l(\vect{x})\,{\rm d}\vect{x} = \delta_{kl}.
\]
Even in the case where the sample points $\{\vect{x}_i\}$ are evenly
spaced, the spacing between them and the extent of the domain ${\cal
D}$ that is sampled are important in determining whether the
corresponding mode vectors $\{\vect{e}_k\}$ are orthonormal. Indeed,
as we shall see below, in some applications it is common for the
corresponding mode vectors {\em not} to be orthonormal. If this is the
case, then it is is no longer true that the dual mode vectors are
identical to the original mode vectors, and so it is incorrect to
calculate the coefficient vector $\hat{\vect{a}}$ using (\ref{standardsoln}),
i.e. by taking the scalar product of the data $\vect{d}$ with the mode
vectors. This will lead to a modal decomposition of the image that is
unnecessarily inaccurate.  Instead, one must return to using the
general result (\ref{dualssoln}), where the mode coefficients are calculated by
taking the scalar product of the data with the corresponding dual mode
vectors. 

Finally, a crucial point is that the converse may also occur. For
example, if a set of continous mode functions are not orthogonal, by
sampling them at a particular set of (non-uniform) points one can
arrive at a set of mode vectors that are orthogonal, in which case the
duals are trivially obtained.

\section{One-dimensional illustrations}
\label{one-dim}

Although the main focus of this paper is the modal decompostion of
two-dimensional astronomical images, it will be informative first to
illustrate the discussion given above by investigating the
decompostion of one-dimensional sampled functions (which may be
considered as cuts through some image). In particular, we will focus
our numerical investigations in this section 
on decompostions using one-dimensional
Gaussian-Hermite (or shapelet) mode functions, as recently advocated
by Refregier (2003) for representing images of galaxies.

\begin{figure}
\begin{center}
\includegraphics[angle=-90,width=0.9\linewidth]{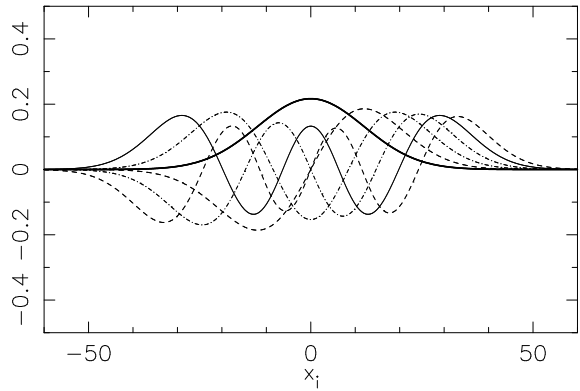}
\caption{The Gaussian-Hermite continuous mode functions
$f_k(x;\beta)$ for $k=0,1,\ldots,5$ and $\beta=12$.}
\label{fig1}
\end{center}
\end{figure}

In this case, the continuous mode functions are given by
\[
f_k(x;\beta) = \beta^{-1/2} \phi_k(\beta^{-1} x),
\]
for $k=0,1,2,\ldots,\infty$. The quantity
$\beta$ is some fixed `characteristic scale' for the set of mode
functions and the function $\phi_k(u)$ reads
\[
\phi_k(u) = (2^k \pi^{1/2} k!)^{-1/2} H_k(u) \exp(-u^2/2),
\]
where $H_k(u)$ is a Hermite polynomial of order $k$. 
From these expressions it is clear that the mode functions are real 
and that $\beta$ is equal to the dispersion `$\sigma$'
of the $k=0$ Gaussian mode function.
It is also 
straightforward to show that the mode functions 
are orthonormal over the domain $-\infty < x < \infty$, i.e.
\begin{equation}
\int_{-\infty}^\infty f_k(x;\beta)f_l(x;\beta)\,{\rm d}x=\delta_{kl}.
\label{orthogdef}
\end{equation}
We note that, for convenience, the mode functions are centred at
$x=0$. It is, of course, possible to centre them about some arbitrary
point $x=x_c$, but in most cases the coordinate system is chosen so
that the centroid of the galaxy image (say) is located at the origin.
The first six mode functions are shown in Fig.~\ref{fig1} for
$\beta=12$.

Let us now consider the decomposition of a 
one-dimensional pixelised `image' into shapelets. 
In any such decomposition, one is free to choose both the 
characteristic scale $\beta$ of the
modes set and the total number $M$ of modes used (note 
that $M=k_{\rm max}+1$). As shown by Refregier (2003), 
such a set of shapelet modes 
is suitable for describing features in an image on scales between
the two limits
\begin{equation}
\theta_{\rm min} \approx \beta M^{-1/2}, \qquad
\theta_{\rm max} \approx \beta M^{1/2}.
\label{scalesdef}
\end{equation}
Thus, if the image has features on scales ranging from $\theta_{\rm
min}$ (e.g. the pixel size or the size of the point spread function) to
$\theta_{\rm max}$ (e.g. the size of the image or the overall extent
of the structure it contains), then a good choice of $\beta$ and
$M$ is
\begin{equation}
\beta \approx (\theta_{\rm min}\theta_{\rm max})^{1/2}, \qquad
M\approx \frac{\theta_{\rm max}}{\theta_{\rm min}}.
\label{criterion1}
\end{equation}

For each of the pixelised test images we consider below, we take the 
coordinate positions of the $N$ (regularly-spaced) sample points to be
\[
\{x_i\}=\{-\tfrac{1}{2}(N-1),\ldots,-1,0,1,\ldots,\tfrac{1}{2}(N-1)\},
\]
thereby giving an image of spatial extent of $N$, with a pixel
size of unity. 
In each case, we take $N=81$.
Taking into account that our test images contain structure on the
order of the image size, but do not contain structure
on scales less than a few pixels, acceptable choices for 
the scale parameter and the number of modes to use are
$\beta \approx 10$ and $M \approx 20$. For convenience, we fix 
$M = 20$, and investigate mode sets for which
$\beta = 1$ -- 20 in steps of unity. In each case, the elements of the 
corresponding pixelised mode vectors are easily calculated from 
$(\vect{e}_k)_i = f_k(x_i;\beta)$. 

\begin{figure}
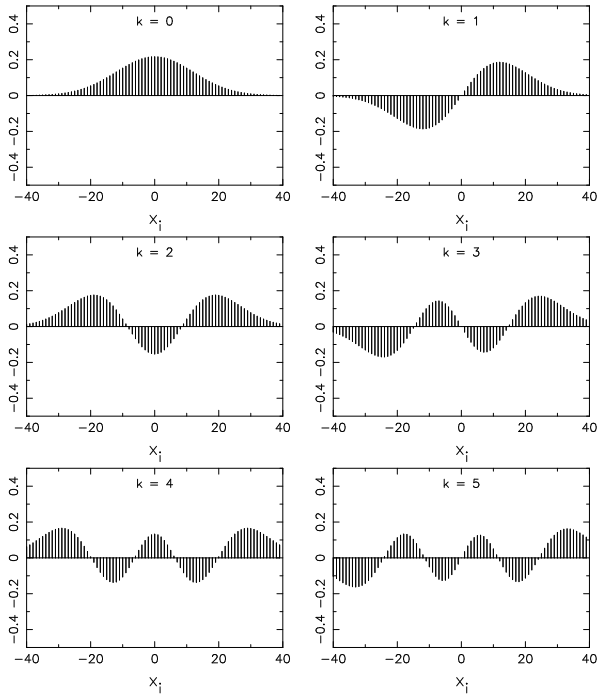

\begin{center}
\includegraphics[angle=-90,width=0.45\linewidth]{herms-0-81-mode.ps}\quad
\includegraphics[angle=-90,width=0.45\linewidth]{herms-1-81-mode.ps} 
\includegraphics[angle=-90,width=0.45\linewidth]{herms-2-81-mode.ps}\quad
\includegraphics[angle=-90,width=0.45\linewidth]{herms-3-81-mode.ps}
\includegraphics[angle=-90,width=0.45\linewidth]{herms-4-81-mode.ps}\quad
\includegraphics[angle=-90,width=0.45\linewidth]{herms-5-81-mode.ps}
\caption{The Gaussian-Hermite mode vectors
$\vect{e}_k$ for $k=0,1,\ldots,5$ and $\beta=12$.}
\label{fig2}
\end{center}
\end{figure}

In Fig.~\ref{fig2} we plot the elements of
the first six mode vectors for $\beta=12$.
We see that, even for this modest value of $\beta$, the support of the
mode vectors for $k>2$ exceeds the length of the image. Indeed, as $k$
increases the support of the mode vectors increases according to
(\ref{scalesdef}), and so this effect becomes extremely pronounced for the
high-$k$ modes. As a result, we would
expect the mode set does not satisfy the discretized form of the
orthogonality condition (\ref{orthogdef}). We will show below that this is
indeed the case.

\begin{figure}
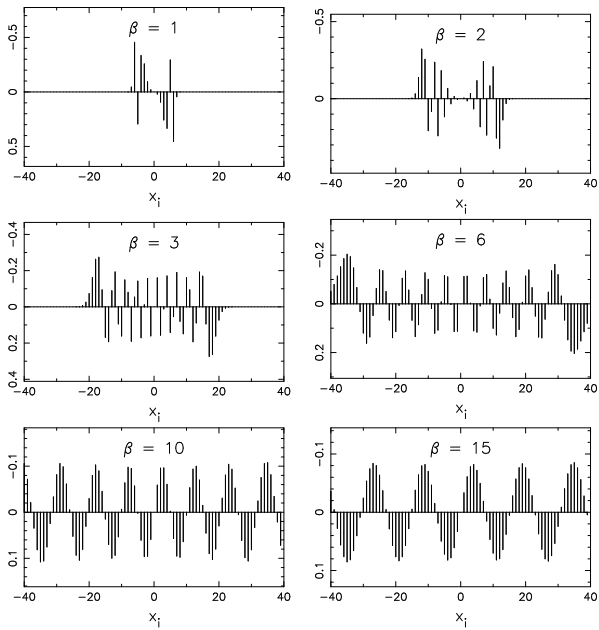

\begin{center}
\includegraphics[angle=-90,width=0.45\linewidth]{1-mode-19.ps}\quad
\includegraphics[angle=-90,width=0.45\linewidth]{2-mode-19.ps} 
\includegraphics[angle=-90,width=0.45\linewidth]{3-mode-19.ps}\quad
\includegraphics[angle=-90,width=0.45\linewidth]{6-mode-19.ps}
\includegraphics[angle=-90,width=0.45\linewidth]{10-mode-19.ps}\quad
\includegraphics[angle=-90,width=0.45\linewidth]{15-mode-19.ps}
\caption{The $k_{\rm max} = 19$ Gaussian-Hermite mode vector
$\vect{e}_{19}$ for $\beta=0,1,2,3,6,10,15$.}
\label{fig3}
\end{center}
\end{figure}

It is also of interest to investigate the structure of the mode
vectors for different values of $\beta$. As an illustration, in
Fig.~\ref{fig3} we plot the highest-$k$ mode vector in the set
($k_{\rm max} = 19$) for six different values of the scale parameter
$\beta$. 
In particular, we note that the mode vector is severely undersampled
for $\beta=1$ and $\beta=2$, in which case the
mode set may again fail to be orthogonal. Moreover, since the
highest-$k$ mode in the set has the largest spatial extent, we see
that, in fact, for $\beta \ge 6$ the support of the mode set exceeds
the length of the image, once again suggesting a loss of
orthogonality.

\subsection{Properties of the mode vectors}
\label{modeproperties}

\begin{figure}
\begin{center}
\includegraphics[angle=-90,width=0.9\linewidth]{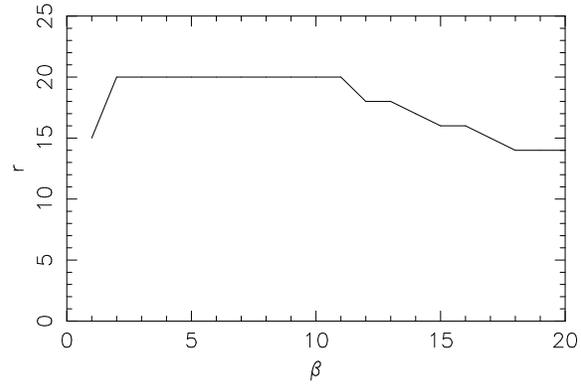}
\caption{The rank $r$ of the mode matrix $\vect{E}$ 
as a function of the scale parameter $\beta$
for $M=20$ Gaussian-Hermite mode vectors.}
\label{fig4}
\end{center}
\end{figure}

Since we have chosen $M<N$, the mode vectors clearly cannot form
a complete basis for the image space. Nevertheless, 
it is of interest to investigate formally the linear-dependence
and orthogonality of the mode vectors $\vect{e}_k$
$(k=0,1,\ldots,M-1)$ for various values of the scale parameter
$\beta$. To this end, for each value of $\beta$ under consideration,
we construct the corresponding mode matrix $\vect{E}$ and then
determine its rank $r$ by calculating its SVD and counting the number
of non-zero singular values. If the mode vectors are linearly
independent, one would expect $r=M$. In Fig.~\ref{fig4}, we plot the
value of $r$ as a function of the parameter $\beta$.
Since the number of mode vectors is $M=20$, we see that the mode
vectors form a linearly-independent $(r=20)$ set only if the scale
parameter lies in the range $\beta = 2$--11. For smaller values of
$\beta$, the undersampling of the underlying continuous mode functions
leads to linear dependence in the resulting mode set. For large 
$\beta$, linear dependence occurs since the support of (some
of) the mode vectors exceeds the length of the image.

\begin{figure}
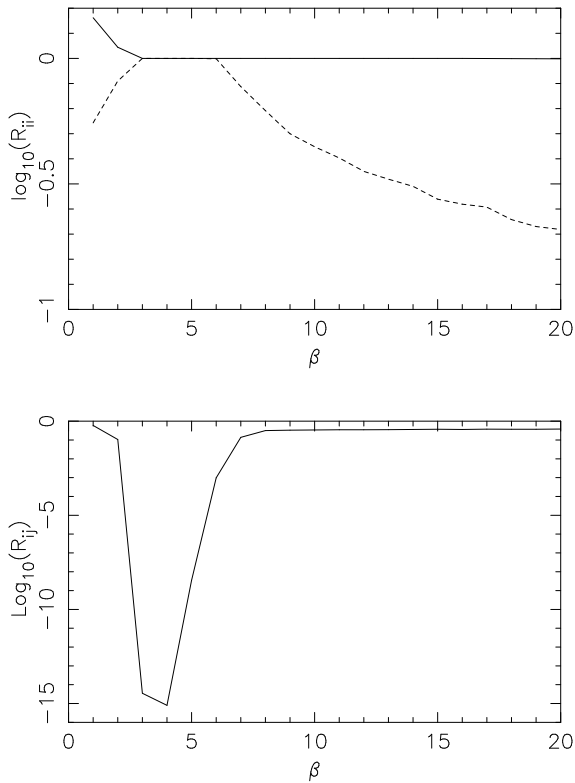

\begin{center}
\includegraphics[angle=-90,width=0.9\linewidth]{Rrank2.ps} \\[0.5cm]
\includegraphics[angle=-90,width=0.9\linewidth]{Rrankoff4.ps}
\caption{Top: the logarithm of the value of the largest (solid line)
and smallest (dashed line) diagonal elements of the Gram matrix
$\vect{R}=\vect{E}^\dagger\vect{E}$ as a function of the scale
parameter $\beta$. Bottom: the logarithm of the largest off-diagonal
element of $\vect{R}$ as a function of $\beta$.}
\label{fig5}
\end{center}
\end{figure}

To investigate whether the mode vectors are orthonormal we calculate
the Gram matrix $\vect{R}=\vect{E}^\dagger\vect{E}$ for each value of
$\beta$ under consideration. Clearly, for an orthonormal set of mode
vectors would expect $\vect{R}=\vect{I}_M$.  In Fig~\ref{fig5} (top), for
each value of $\beta$, we plot the logarithms of 
the largest and smallest diagonal elements of $\vect{R}$.  For mode
vectors of unit length, we would expect all the diagonal elements to
equal unity, and so the two plotted curves should coincide at
$\log(R_{ii})=0$. We see that this occurs only if the scale parameter
lies in the range $\beta = 3$--6.
In the bottom panel of Fig.~\ref{fig5}, we plot the value of the
logarithm of the largest off-diagonal element of $\vect{R}$ as a
function of $\beta$. For an orthogonal set of mode vectors, one would
expect all the off-diagonal elements of $\vect{R}$ to be zero in the
ideal case, or below the machine precision in a numerical
implementation. We see that this is only the case if $\beta = 3$--4,
although the off-diagonal elements remain reasonably small for 
$\beta = 5$--6. For other values of $\beta$, however, we see that
there exist off diagonal elements with absolute values greater than
$\sim 0.1$, which is a clear indication that the corresponding mode
sets are not orthogonal. We also note that there exist values of
$\beta$ for which the modes are non-orthogonal, but still have full rank.

\subsection{Dual mode vectors}
\label{dualmodevectors}

\begin{figure}
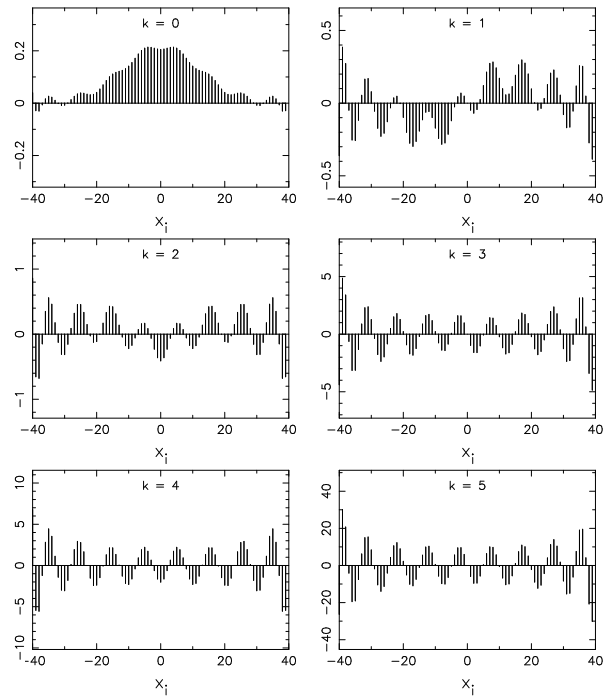

\begin{center}
\includegraphics[angle=-90,width=0.45\linewidth]{duals-12-0-81-mode.ps}\quad
\includegraphics[angle=-90,width=0.45\linewidth]{duals-12-1-81-mode.ps}
\includegraphics[angle=-90,width=0.45\linewidth]{duals-12-2-81-mode.ps}\quad
\includegraphics[angle=-90,width=0.45\linewidth]{duals-12-3-81-mode.ps}
\includegraphics[angle=-90,width=0.45\linewidth]{duals-12-4-81-mode.ps}\quad
\includegraphics[angle=-90,width=0.45\linewidth]{duals-12-5-81-mode.ps}
\caption{The Gaussian-Hermite dual mode vectors
$\tilde{\vect{e}}_k$ for $k=0,1,\ldots,5$ and $\beta=12$.}
\label{fig6}
\end{center}
\end{figure}

Since for mode sets with $\beta < 3$ or $\beta \ga 6$ the mode vectors are not
orthogonal, the corresponding dual mode vectors are not simply
some multiple of the original modes. It is therefore of interest to 
investigate the form of the dual modes in these cases. As an
illustration, in Fig.~\ref{fig6} we plot the first six
dual mode vectors for the case $\beta=12$.
These should be compared with the corresponding original mode vectors
plotted in Fig.~\ref{fig1}.  We see that the form of each dual is very
different from the corresponding original mode vector. In particular,
we note that the form of the duals is determined by the entire
original mode set. As a result, the 
large spatial extent of the large-$k$ original
mode vectors has an effect on the form of the low-$k$ dual mode
vectors, even though the support of corresponding low-$k$ original mode vectors
does not exceed the length of the image. Thus, for example, the $k=0$
and $k=1$ dual vectors are markedly different from the 
correspondinig original mode vectors, even though the latter lie
completely within the image length.

Although clear from the discussion in Section~\ref{dualmodes}, it is
worth pointing out once more that the dual modes are {\em not} the
basis set in terms of which an image is described. In the approach
presented, the image is still represented as a linear combination of
the original Gauss-Hermite (or shapelet) mode vectors. It is simply
that the value of the coefficient $a_k$ in the linear combination is
given by the scalar product of the image vector $\vect{d}$ with the
$k$th dual mode $\tilde{\vect{e}}_k$, rather than with the original
mode vector $\vect{e}_k$. Thus, the advantageous properties of the
Gauss-Hermite modes (or any other mode set under consideration) can
still be used in the analysis of the resulting modal
decomposition. From Fig.\ref{fig6}, however, it is clear that the dual
vectors may not possess the same localisation as the basis vector to
which they relate.  Consequently, for the $k=0$ mode for example, a
feature towards the edge of the field will affect the coefficients of
the mode even though the original mode vector falls to zero there.

\subsection{Decomposition of simple functions}
\label{simpledecomp}

Now that we have discussed the properties of the mode vectors and
their duals for various values of the scale parameter $\beta$, it is
of interest to investigate the effect of taking proper account of
linear dependence and non-orthogonal. To this end, we perform
the modal decomposition of some simple one-dimensional test functions
using both the method advocated in Section~\ref{SVD} and the standard
approach in which the coefficients in the decompostion are obtained
simply by taking scalar products of the image vector with the original
mode vectors (see e.g. Refregier 2003). 
The test functions considered, although simple, are
relevant to the decomposition of astronomical images.

\subsubsection*{Uniform function}

\begin{figure}
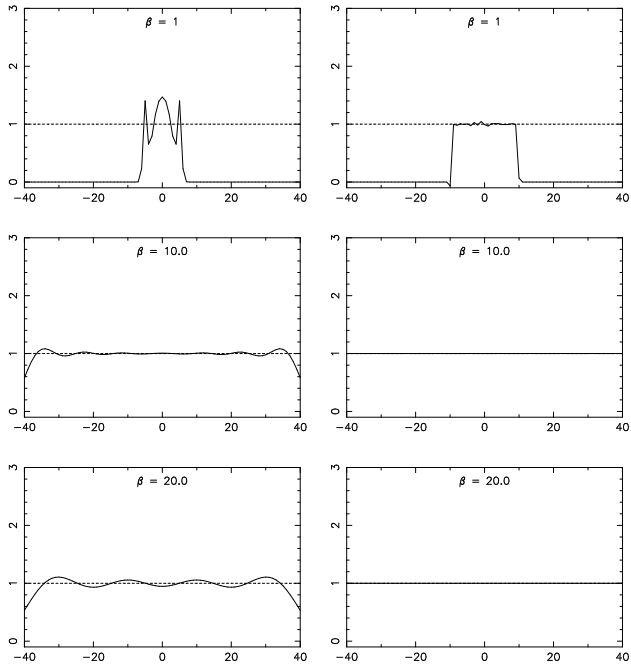

\begin{center}
\includegraphics[angle=-90,width=0.475\linewidth]{uni-comp-20-41-T-1.ps}\quad
\includegraphics[angle=-90,width=0.475\linewidth]{uni-comp-20-41-E-1.ps} 
\includegraphics[angle=-90,width=0.475\linewidth]{uni-comp-20-41-T-10.0.ps}\quad
\includegraphics[angle=-90,width=0.475\linewidth]{uni-comp-20-41-E-10.0.ps}
\includegraphics[angle=-90,width=0.475\linewidth]{uni-comp-20-41-T-20.0.ps}\quad
\includegraphics[angle=-90,width=0.475\linewidth]{uni-comp-20-41-E-20.0.ps}
\caption{The uniform function (dashed line) and its modal
  decomposition (solid line) into $M=20$ Gaussian-Hermite mode vectors
  with scale parameter $\beta$. The decomposition coefficients are
  obtained using the standard method (left column) and the SVD
  method (right column).}
\label{fig7}
\end{center}
\end{figure}

We begin by considering the modal decomposition of a uniform
signal. The issue of accommodating (nearly) uniform background
emission is clearly relevant in an astronomical context. 
The modal decomposition of this function was calculated, 
using both the standard method and the SVD method, for $M=20$ modes
and for integer
values of the scale parameter in the range $\beta = 1$--20.
In Fig.~\ref{fig7} we plot the resulting decompositions for the
standard method (left column) and the approach advocated here (right
column) for some representative values of $\beta$.
\begin{figure}
\begin{center}
\includegraphics[angle=-90,width=0.8\linewidth]{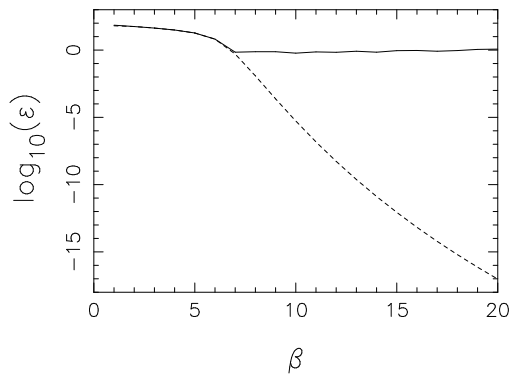}
\caption{Decomposition residual $\epsilon = |\vect{Ea}-\vect{d}|$ as a
  function of $\beta$ for the
  modal decomposition of the uniform function using the standard method
 (solid line) and the SVD method (dashed line).}
\label{fig8}
\end{center}
\end{figure}
These values have been chosen to correspond to mode sets for which the
mode vectors are linearly dependent and non-orthogonal, as
demonstrated in Fig.~\ref{fig5}. In Fig.~\ref{fig8}, we plot the
residual $\epsilon = |\vect{Ea}-\vect{d}|$ of the decompostions as a
function of $\beta$ for the standard method (solid line) and the SVD
method (dashed line).

We see from Fig.~\ref{fig7} that, as one
would expect, the decompositions produced by the two methods differ,
in some cases significantly. It was confirmed, however, that for mode
sets with $\beta=3$ and $\beta=4$, for which the mode vectors are
linearly-independent and orthonormal, the modal decompositions
produced by the two methods coincide to machine precision.
In the cases illustrated in Fig.~\ref{fig7}, we see that the SVD
method provides a decomposition that is consistently better than that
of obtained with the standard approach.  In particular, we note that
for a wide range of $\beta$-values, the SVD method produces a
decomposition that is visually indistinguishable from the input
function, whereas the standard approach exhibits a pronounced ringing
for all values of $\beta$. A quantitative description of the improved
decomposition quality is given by Fig.~\ref{fig8}. As expected, the
SVD approach always produces the smallest decomposition residuals,
indeed by many orders of magnitude for $\beta \ga 7$. 
Since the decomposition
of an image is a linear process, the ability to describe
accurately a uniform function enables better decompositions of
discrete objects in a uniform background, which is clearly important
in astronomical applications.
We also note
that, for the SVD method, the decomposition residual is 
a monotonically decreasing function of the scale parameter $\beta$.
For the standard method, however, there is a shallow minimum in the
residual at $\beta=7$. It is interesting that, from Fig.~\ref{fig5},
for this value of $\beta$ the mode set is neither linearly-independent
nor orthogonal.

\subsubsection*{Top-hat function}

\begin{figure}
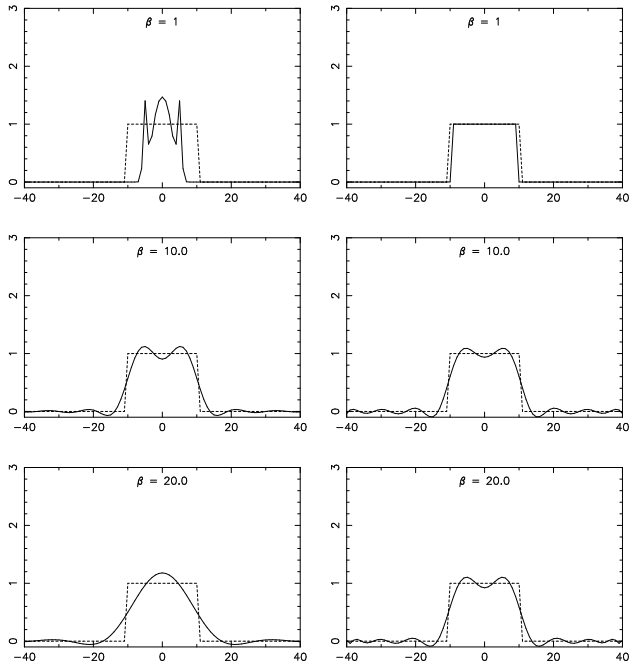

\begin{center}
\includegraphics[angle=-90,width=0.475\linewidth]{th-comp-20-41-T-1.ps}\quad
\includegraphics[angle=-90,width=0.475\linewidth]{th-comp-20-41-E-1.ps} 
\includegraphics[angle=-90,width=0.475\linewidth]{th-comp-20-41-T-10.0.ps}\quad
\includegraphics[angle=-90,width=0.475\linewidth]{th-comp-20-41-E-10.0.ps}
\includegraphics[angle=-90,width=0.475\linewidth]{th-comp-20-41-T-20.0.ps}\quad
\includegraphics[angle=-90,width=0.475\linewidth]{th-comp-20-41-E-20.0.ps}
\caption{As for Fig.~\ref{fig7}, but for the top-hat function of width
  20.}
\label{fig9}
\end{center}
\end{figure}

\begin{figure}
\begin{center}
\includegraphics[angle=-90,width=0.8\linewidth]{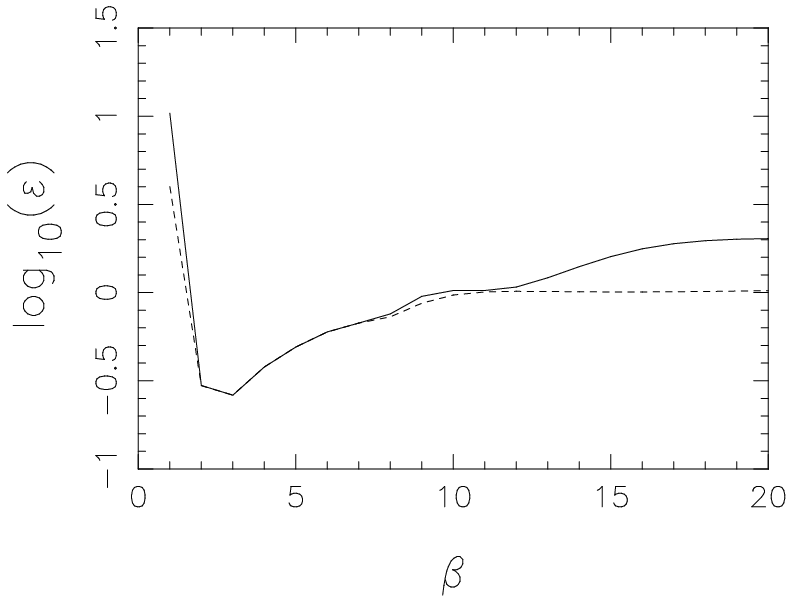}
\caption{As in Fig.~\ref{fig8}, but for the top-hat function of width 20.}
\label{fig10}
\end{center}
\end{figure}

As our second illustration, we consider a top-hat function of width
equal to 20 units. Once again, this simple function is relevant to
astronomical image analysis of, for example, saturated images of
compact objects. The modal decomposition of this function was 
again calculated, 
using both the standard method and the SVD method, for integer
values of the scale parameter in the range $\beta = 1$--20.
In Fig.~\ref{fig9} we plot the resulting decompositions for the
standard method (left column) and the SVD method (right
column) for some representative values of $\beta$.
The decomposition residuals are plotted in Fig.~\ref{fig10} 
as a function $\beta$ for the standard method (solid line) and the 
method (dashed line).

We see from Fig.~\ref{fig9} that, once again, the two methods produce
different decompositions for the values of $\beta$ illustrated, and
that the SVD method produces superior results. For both methods,
however, the presence of only $M=20$ modes leads to considerable
ringing in the decompostions for almost all values of $\beta$.
Nevertheless, it is interesting that for $\beta=1$ the standard method
is unable to produce a reasonable reconstruction, whereas an excellent
decomposition is obtained over a limited extent of the function using
the SVD method, the extent corresponding to the span of the mode
set. The quantitative comparison of the two methods given in
Fig.~\ref{fig10} once again shows that, by design, the SVD method
always produces the most accurate decomposition. In this case,
however, the difference between the two methods is not as great as it
was for the decomposition of the uniform function. We also note that
the residuals for both methods in this case are minimised at $\beta =
3$. At this point, the minimum residual is identical for the two
methods, since the original mode set is linear-independent and
orthonormal to machine precision.

\subsubsection*{$\beta$-model profile}

Our final one-dimensional illustration is that of a $\beta$-model
profile. Clusters of galaxies are often modelled (King 1972) as
spherically-symmetric, with an electron density profile of the form
\[
n_{\rm e}(r) =
n_0\left[1+\left(\frac{r}{r_c}\right)^2\right]^{-3b/2},
\]
where $r_c$ is the cluster core radius and $b$ is a constant (this is usually
denoted by $\beta$, but this has already been used in this paper to denote
the scale parameter of the mode set).  
In this case, one easily finds that the projected
thermal Sunyaev-Zel'dovich profile of the cluster is
given by
\begin{equation}
f(x)=f_0\left(1+\frac{x^2}{x_c^2}\right)^{-\lambda},
\label{king}
\end{equation}
where $x$ the angular separation on the sky and $\lambda=(3b-1)/2$.
Assuming the standard value $b=2/3$ gives $\lambda=1/2$, and we use
this value for our test function, together with a core length $x_c = 3$.

The modal decomposition of this function was calculated, 
using both the standard method and the SVD method, for integer
values of the scale parameter in the range $\beta = 1$--20.
In Fig.~\ref{fig11} we plot the resulting decompositions for the
standard method (left column) and the SVD method (right
column) for some representative values of $\beta$.
\begin{figure}
\begin{center}
\includegraphics[angle=-90,width=0.475\linewidth]{cauch-comp-20-41-T-1.ps}\quad
\includegraphics[angle=-90,width=0.475\linewidth]{cauch-comp-20-41-E-1.ps} 
\includegraphics[angle=-90,width=0.475\linewidth]{cauch-comp-20-41-T-10.ps}\quad
\includegraphics[angle=-90,width=0.475\linewidth]{cauch-comp-20-41-E-10.ps}
\includegraphics[angle=-90,width=0.475\linewidth]{cauch-comp-20-41-T-20.ps}\quad
\includegraphics[angle=-90,width=0.475\linewidth]{cauch-comp-20-41-E-20.ps}
\caption{As for Fig.~\ref{fig7}, but for the $\beta$-model profile with
  core radius $x_c=3$}
\label{fig11}
\end{center}
\end{figure}
\begin{figure}
\begin{center}
\includegraphics[angle=-90,width=0.8\linewidth]{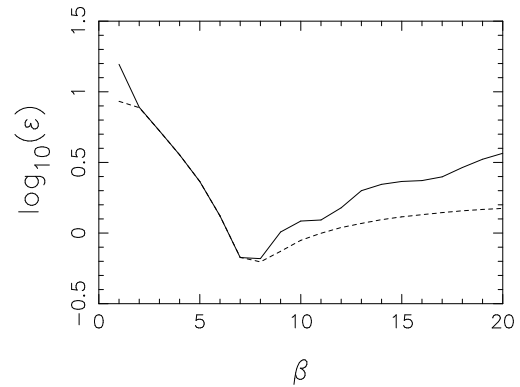}
\caption{As for Fig.~\ref{fig8}, but for the $\beta$-model profile
  with core radius $x_c=3$.}
\label{fig12}
\end{center}
\end{figure}
The decomposition residuals are plotted in Fig.~\ref{fig12} 
as a function $\beta$ for the standard method (solid line) and the SVD
method (dashed line).

From Fig.~\ref{fig11}, we see again that the SVD methods produces
superior decompositions for the values of $\beta$ illustrated.
For both methods, some ringing is observed in the decompositions, but
this is much less pronounced for the SVD method. In particular, for
large values of $\beta$, the standard method cannot accurately
reproduce the narrow peak of the $\beta$-model, whereas this is
achieved by the SVD method. We also note that, for $\beta=1$, the
SVD method reproduces the central sharp peak very accurately.
The quantitative comparison of the two methods given in
Fig.~\ref{fig11} clearly shows the improvement obtained using the
SVD method, which produces a residual significantly below
that obtained using the standard method for $\beta \ga 8$. This is also
the value of the scale parameter for which both methods yield
their lowest residual, and the corresponding set of
original mode vectors are close to linearly-dependent and non-orthogonal.

\section{Image decompostion}
\label{imagedecomp}
We now consider the modal decomposition of real two-dimensional
astronomical images.  These images were obtained by first using ${\tt
SExtractor}$ (Bertins \& Arnouts 1996) on the Hubble Deep Fields
(HDFs; Williams et al. 1996, 1998). The convolution mask and detection
parameters were adapted from those used by Massey et al. (2003). In
particular, in order to allow the recovery of faint objects, we
adopted a low signal-to-noise detection threshold, {\sc
detect\_thresh}, of 1.3. Objects with {\sc class\_star} $>$ 97 per
cent were discarded, since we wish to analyse only galaxies. The image
was then segmented into small square `postage stamp' regions around
the remaining galaxies. The size of the images were set to $51 \times
51$ pixels. In this section, we investigate the decomposition of the
HDF galaxy images into a set of two-dimensional Gaussian-Laguerre
modes.


\subsection{Gauss-Laguerre mode functions}

The two-dimensional Gauss-Laguerre continuous mode functions are the
simultaneous eigenstates of energy and angular momentum for the
two-dimensional isotropic quantum harmonic oscillator. Interestingly,
these mode functions are also the solutions in polar coordinates to
the paraxial wave equation in optics (see e.g. Goldsmith 1998). The
relationship between the standard forms of these mode functions used
in the above contexts is discussed in detail in the Appendix.

In the field of optical and quasi-optical systems, it is customary
(Murphy et al. 1996; Withington et al. 2000) to define the set of
mode functions as
\begin{equation}
\psi_p^m(r,\theta) = \left[\frac{(2-\delta_{0m})p!}{\pi
    \beta^2 (p+|m|)!}\right]^{1/2} x^{|m|/2} \mbox{e}^{-x/2} 
L_p^{|m|}(x)\,\mbox{e}^{\mbox{i}m\theta},
\label{pm-decomp}
\end{equation}
where $x=r^2/\beta^2$ and $L_p^{|m|}(x)$ are the standard associated
Laguerre polynomials.  The functions $\psi_p^m$ are conventionally
called the Gaussian beam $pm$-modes or simply the $pm$-modes (see
e.g. Goldsmith 1998) and form an orthonormal set over the infinite
Euclidean plane.  The radial index can take the values $p =
0,1,2,\ldots,\infty$, and the azimuthal index $m$ can take the values
$m=0,\pm 1, \pm 2, \ldots, \pm \infty$.  In practice, it is customary
to truncate the mode set by imposing maximum values for $p$ and
$|m|$. Indeed , it is usual to choose $|m|_{\rm max} = p_{\rm max}$,
which leads to a total number of modes ${\cal N}_1 = (p_{\rm
max}+1)(2p_{\rm max}+1)$.  If $f(r,\theta)$ is real then clearly
$a_p^{-m} = (a_p^m)^*$.

In the context of quantum mechanics, it is more usual to label the
modes in terms of the energy quantum number $n=2p+|m|$ and $m$, rather
than $p$ and $m$. Indeed, Refregier (2003) uses this convention to
arrive at the polar shapelet modes
\begin{equation}
\phi_n^m(r,\theta) = \left[\frac{(2-\delta_{0m})(\tfrac{n-|m|}{2})!}{\pi
    \beta^2 (\tfrac{n+|m|}{2})!}\right]^{1/2} x^{|m|/2} \mbox{e}^{-x/2}
L_{\frac{n-|m|}{2}}^{|m|}(x) \, \mbox{e}^{\mbox{i}m\theta},
\label{nm-decomp}
\end{equation}
which also form an orthonormal set over the infinite Euclidean plane.
\footnote{We note that this definition of the polar shapelet functions
differs somewhat from that given in Refregier (2003) and Massey et
al. (2003), which contain some typographical errors.} The energy
quantum number can take the values $n=0,1,2,\ldots,\infty$, and the
azimuthal (angular momentum) quantum number takes the values $m = -n,
-n+2, \ldots, n-2, n$, thereby giving $n+1$ values of $m$ for each
value of $n$.  In practice, one truncates the mode set by imposing a
maximum value for $n$, which leads to a total number of modes ${\cal
N}_2 = \frac{1}{2}(n_{\rm max}+1)(n_{\rm max}+2)$.  Once again, if
$f(r,\theta)$ is real, it is clear that $b_p^{-m} = (b_p^m)^*$.

\subsection{Properties of the mode sets}

It is clear from the above discussion (and that given in the Appendix)
that the two decompositions (\ref{pm-decomp}) and (\ref{nm-decomp})
use very different combinations of Gauss-Laguerre modes. The question
thus arises as to the relative merits of the two mode sets in
decomposing a two-dimensional astronomical image. It is therefore of
interest to investigate the properties of pixelised versions of these
two mode sets.

As shown by Refregier (2003), by analogy with the criteria
(\ref{criterion1}), the appropriate values of the scale parameter
$\beta$ and the maximum energy quantum number $n_{\rm max}$ for the
polar shapelet mode set (\ref{nm-decomp}) are given by
\begin{equation}
\beta \approx (\theta_{\rm min}\theta_{\rm max})^{1/2}, \qquad
n_{\rm max} \approx \frac{\theta_{\rm max}}{\theta_{\rm min}}-1
\label{2dcrit}
\end{equation}
for the decomposition of an image with features on scales ranging from
$\theta_{\rm min}$ to $\theta_{\rm max}$.  The galaxy images to be
decomposed are of size $51 \times 51$ pixels, and contain structure
down to scales of around 3 pixels. Using the above criteria, we
therefore choose $n_{\rm max}=18$.  Thus, the total number of polar
shapelet modes is 190. To allow a fair comparison between the polar
shapelet and $pm$-mode sets, one should arrange for the two sets to
contain (as close as possible) the same total number of
modes. Fortunately, by choosing $p_{\rm max} = 9$, one can arrange for
the $pm$-mode set also to contain $190$ modes.  The corresponding
$pm$-mode and polar shapelet mode vectors are obtained by pixelising
the continuous mode functions on the same grid as the galaxy
images. Each mode set contains only $190$ modes, which is clearly far
less than the number of pixels in the image, but need not be less than
the number of degrees of freedom in some class of images. For an
arbitrary image of infinite resolution the number of degrees of
freedom and pixels will be the same and both decompositions can only
represent an approximation to the image.
\begin{figure}
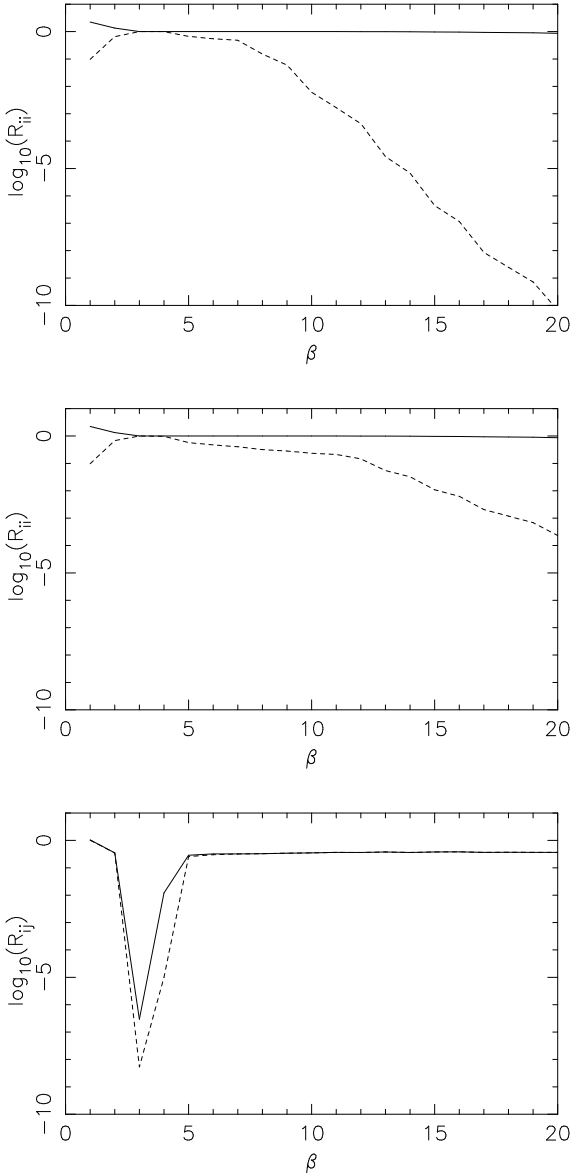

\begin{center}
\includegraphics[angle=-90,width=0.9\linewidth]{Rrank2fp.ps} \\[0.5cm]
\includegraphics[angle=-90,width=0.9\linewidth]{Rrank2ft.ps} \\[0.5cm]
\includegraphics[angle=-90,width=0.9\linewidth]{Rrank2foff.ps}
\caption{Top: the logarithm of the value of the largest (solid line)
and smallest (dashed line) diagonal elements of the Gram matrix
$\vect{R}=\vect{E}^\dagger\vect{E}$ as a function of the scale
parameter $\beta$ for the polar shapelet mode set. Middle: the same,
but for the $pm$-mode set. Bottom: the logarithm of the largest off-diagonal
element of $\vect{R}$ as a function of $\beta$ for the polar shapelets
(dashed line) and $pm$-mode set (solid line).}
\label{fig13}
\end{center}
\end{figure}

We may investigate the properties of the polar shapelet and $pm$-mode
sets in the same manner as we analysed the one-dimensional shapelet
modes. To this end, for each integer value of $\beta$ in the range 0--20,
we construct the mode matrix $\vect{E}$ for both the polar shapelets
and $pm$-modes. The rank $r$ of both matrices was found to equal the
number of modes (190) for all values of $\beta$ under consideration,
hence showing that both modes sets are linearly-independent in all
cases. To investigate whether the mode vectors in each set are
orthonormal, we calculate the Gram matrix $\vect{R}=\vect{E}^\dagger
\vect{E}$ for each mode set for each value of $\beta$. For an
orthonormal mode set $\vect{R}$ should equal the identity matrix. 
In Fig.~\ref{fig13} (top and middle), we plot the logarithm of the
value of the largest and smallest diagonal element of $\vect{R}$ for
the two mode sets. We see that, for both mode sets, the two curves
coincide at $\log(R_{ii})=0$ only for $\beta = 3$ and 4, and so it is
only for these values of the scale parameter that the mode vectors are
of unit length. In the bottom panel of Fig.~\ref{fig13}, we plot the
logarithm of the largest off-diagonal element of $\vect{R}$ for the
polar shapelet (dashed line) and $pm$-mode set (solid line). We see
that, for each mode set, the largest off-diagonal element is only
below the machine precision for $\beta=3$, and so it is only for this
case that either mode set is orthogonal.

\subsection{Decomposition of HDF galaxy images}

For the galaxy images to be decomposed, the result (\ref{2dcrit})
suggests one should set $\beta=14$ for the polar shapelets; for
comparison purposes we will also use this value of the scale parameter
for the $pm$-mode set. 

\begin{figure*}
\begin{center}
\includegraphics[width=0.25\linewidth]{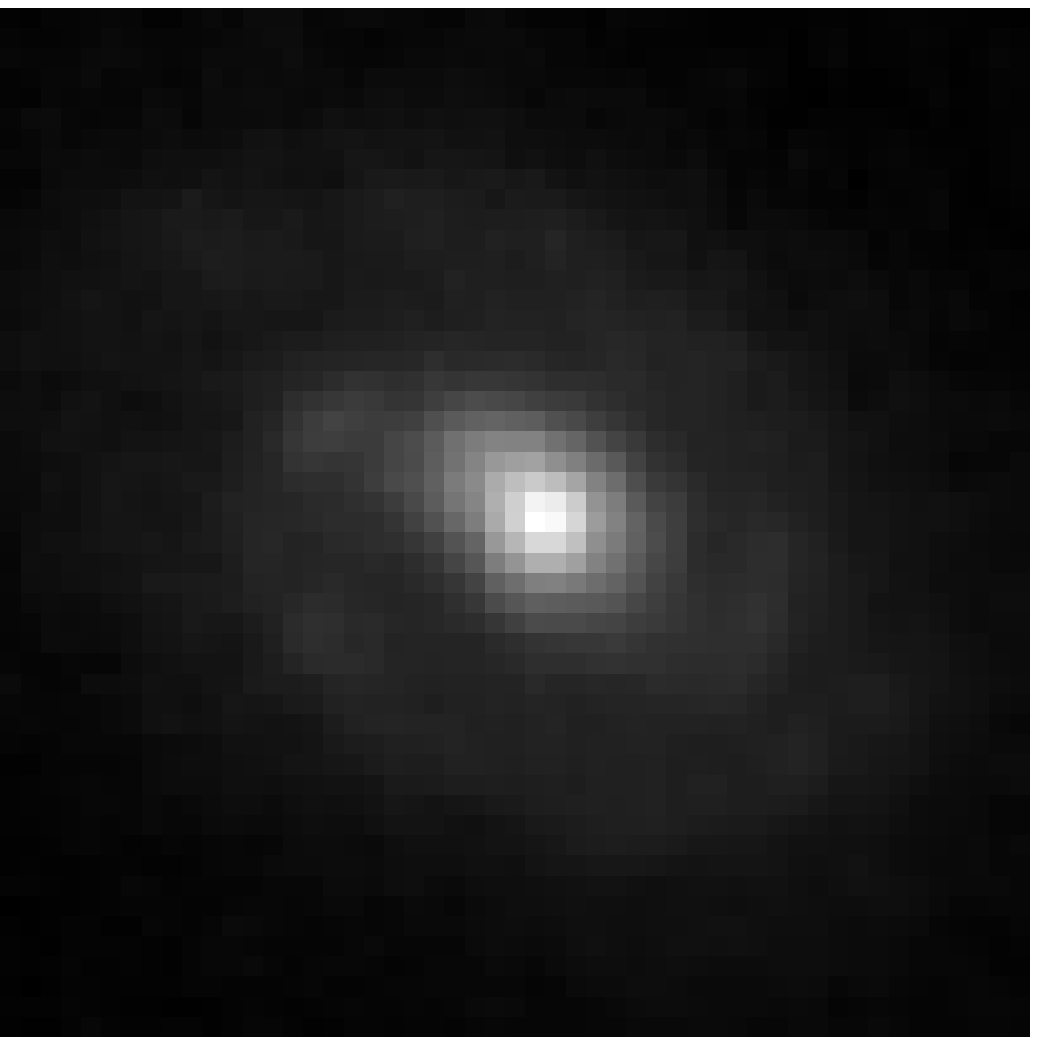}\quad
\includegraphics[width=0.35\linewidth]{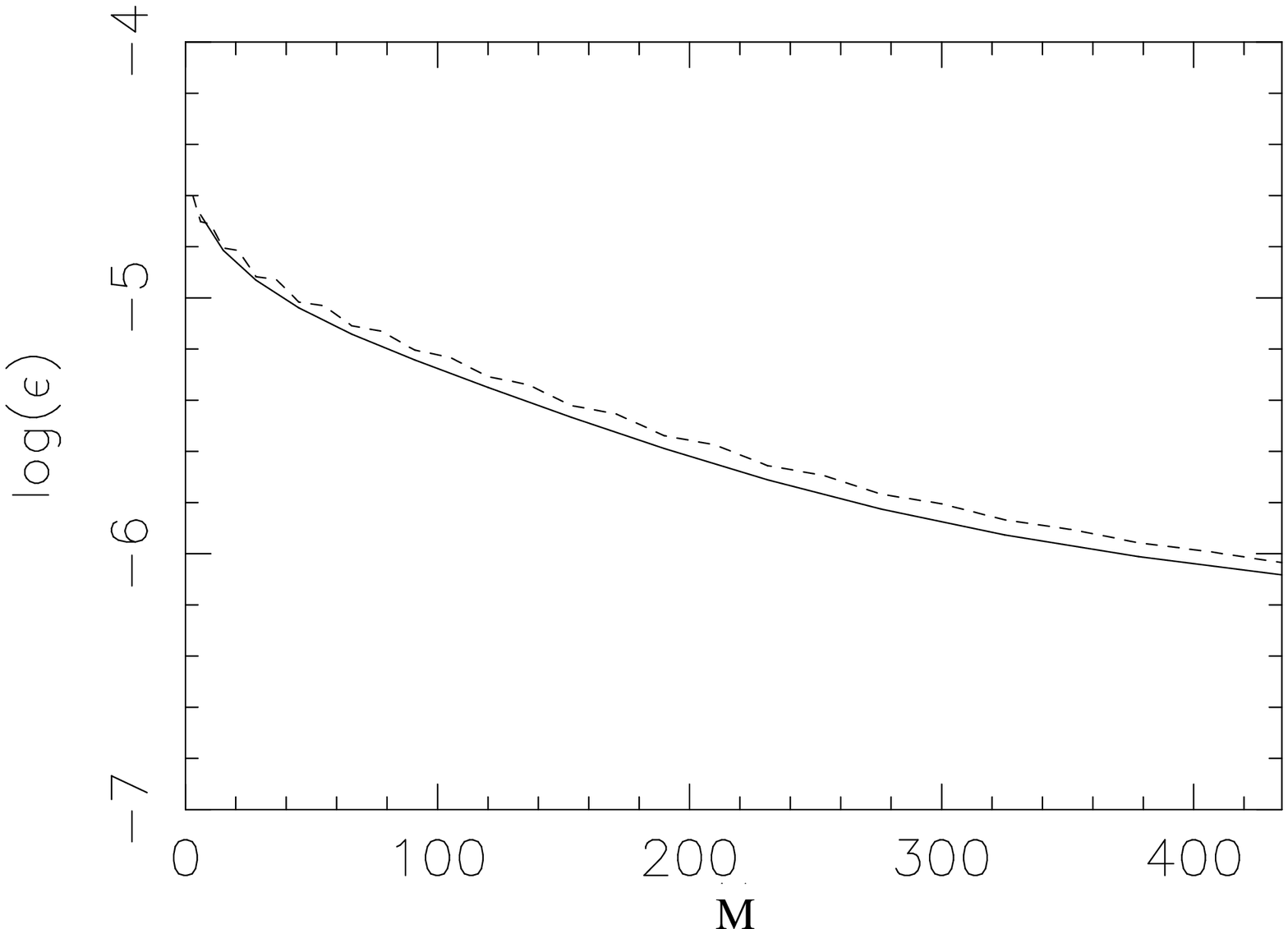}
\includegraphics[width=0.35\linewidth]{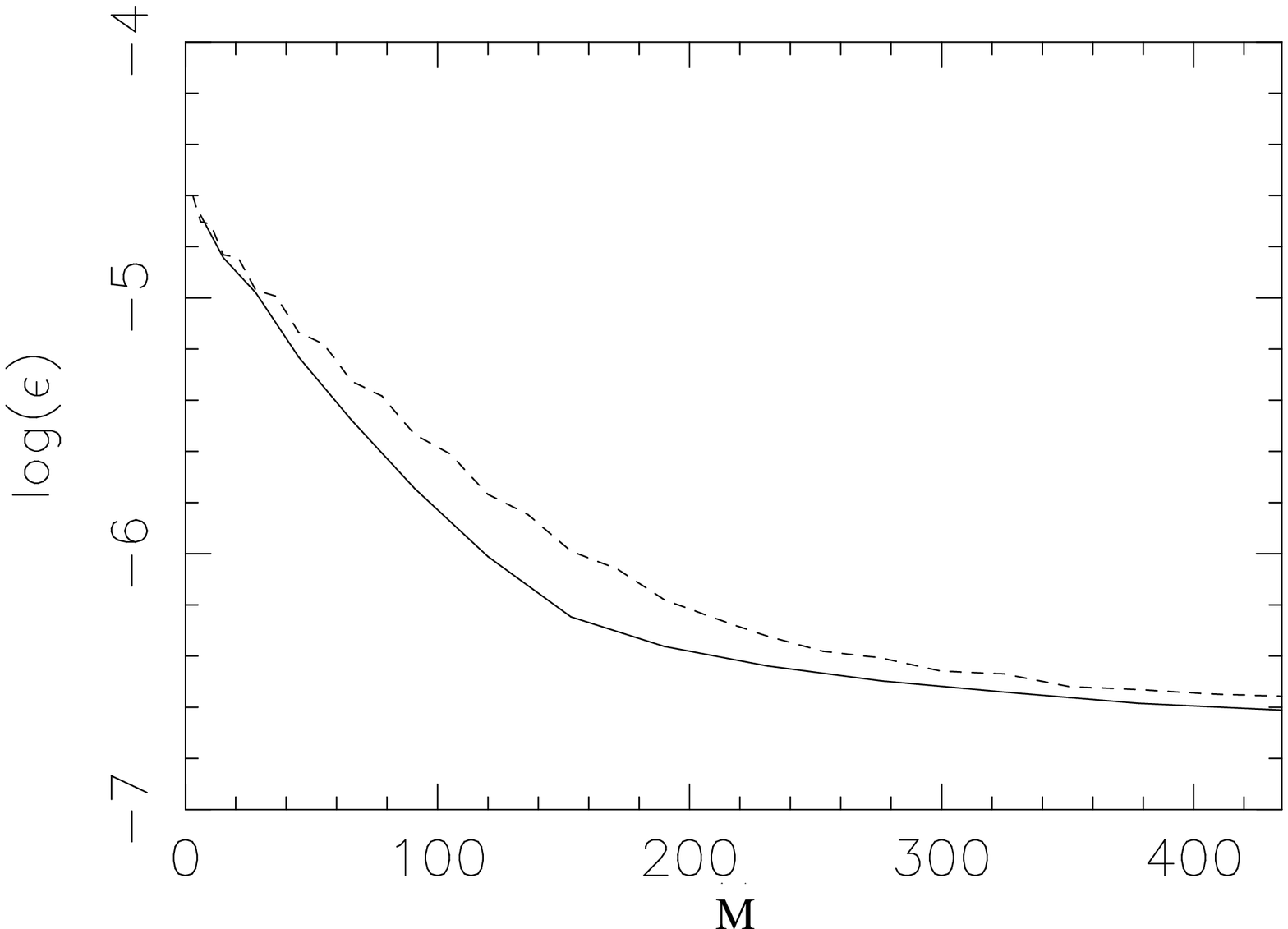}
\caption{Left: a Hubble Deep Field galaxy. 
Middle: the decomposition residual using the standard approach for the
polar shapelets (dashed line) and the $pm$-mode set (solid line) as a
function of the total number of modes $M$ in the set. 
Right: as middle panel, but using the SVD approach.}
\label{fig14}
\end{center}
\end{figure*}

To investigate the relative merits of the
two mode sets, we first consider the decomposition of the single
galaxy image shown in Fig.~\ref{fig14} (left). In the middle panel, we
plot the decomposition residual using the standard approach for the
polar shapelets (dashed line) and the $pm$-modes (solid line) as a
function of the total number of modes $M$ in the set. As expected, for
both mode sets, the decomposition residual decreases as more modes are
used. We also see, however, that, for the same total number of modes,
the $pm$-modes yield a slightly lower decomposition residual than the
polar shapelets. In the right-hand panel of Fig.\ref{fig14}, plot the
corresponding results obtained using the SVD appraoch. It is clear
that, for any given value of $M$, the SVD approach outperforms the
standard approach. We also see that, once again, the $pm$-modes yield
a lower decomposition residual than the polar shapelets.

\begin{figure}
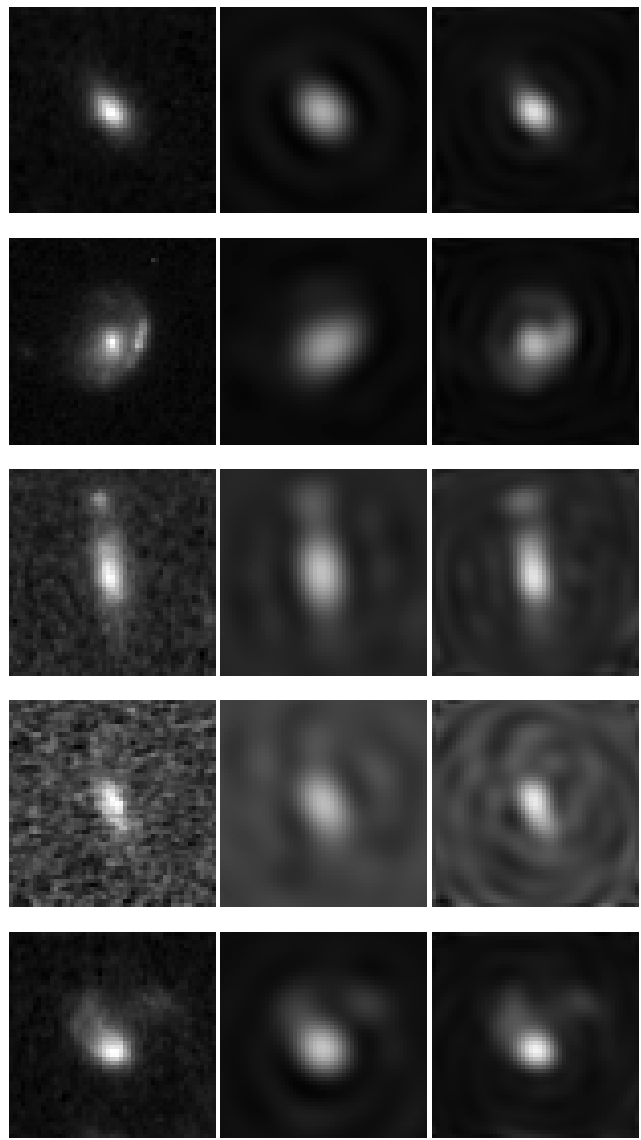

\begin{center}
\includegraphics[width=0.325\linewidth]{original0199-10-16.9705627485.ps}
\includegraphics[width=0.325\linewidth]{T-Rec-0199-10-16.9705627485.ps}
\includegraphics[width=0.325\linewidth]{E-Rec-0199-10-16.9705627485.ps}\\[3mm]
\includegraphics[width=0.325\linewidth]{original0138-10-16.9705627485.ps}
\includegraphics[width=0.325\linewidth]{T-Rec-0138-10-16.9705627485.ps}
\includegraphics[width=0.325\linewidth]{E-Rec-0138-10-16.9705627485.ps}\\[3mm]
\includegraphics[width=0.325\linewidth]{original0139-10-16.9705627485.ps}
\includegraphics[width=0.325\linewidth]{T-Rec-0139-10-16.9705627485.ps}
\includegraphics[width=0.325\linewidth]{E-Rec-0139-10-16.9705627485.ps}\\[3mm]
\includegraphics[width=0.325\linewidth]{original0146-10-16.9705627485.ps}
\includegraphics[width=0.325\linewidth]{T-Rec-0146-10-16.9705627485.ps}
\includegraphics[width=0.325\linewidth]{E-Rec-0146-10-16.9705627485.ps}\\[3mm]
\includegraphics[width=0.325\linewidth]{original0183-10-16.9705627485.ps}
\includegraphics[width=0.325\linewidth]{T-Rec-0183-10-16.9705627485.ps}
\includegraphics[width=0.325\linewidth]{E-Rec-0183-10-16.9705627485.ps}
\caption{Hubble Deep Field galaxies (left column) decomposed into polar
  Gauss-Laguerre $pm$-modes with $p_{\rm max}=9$ and $\beta=14$ using
  the standard method (middle column) and the SVD method (right column).}
\label{fig15}
\end{center}
\end{figure}

The decomposition of five representative galaxy images 
into the $pm$-mode set are
shown in Fig.~\ref{fig15}, using both the standard technique (middle
column) and the SVD method (right column).  As anticipated, since the
mode set is not orthogonal, the two decompositions differ, markedly in
some cases. For each galaxy image, however, we see that the SVD method
produces a more faithful reconstruction. In particular, we note that
the SVD method is more successful in reproducing the finer detail of
the image. By contrast, the decompositions obtained using the standard
approach contain structure only on larger scales. A quantitative
comparison of the decomposition quality is given in Table~\ref{tab1},
which lists the values of the residual $\epsilon$ of the
decomposition. As expected, in each case the SVD method produces a
more accurate decomposition than the standard method.

\section{Discussion and conclusions}
\label{conc}

We have discussed a general method for performing optimal image
decomposition into a set of arbitrary digitised mode functions, which
is an everyday problem in astronomy. Our approach may be applied
straightforwardly even if the modes are linearly-dependent,
non-orthogonal or incomplete, although in the last case the
decomposition can only, in general, approximate the original image.
By constructing the matrix $\vect{E}$ containing the digitised modes,
the properties of the mode set can be easily obtained by performing a
singular value decomposition (SVD). Moreover, having obtained the SVD,
the optimal values for the coefficients in the linear decomposition
can be obtained immediately. If desired, the optimal coefficients may
also be obtained by constructing the set of dual modes, and
performing scalar products of these duals with the input data vector.
This should be contrasted with the standard method of taking scalar
products of the original mode vector with the data vector. This latter
approach yields optimal results only in the case where the digitised
mode vectors form an orthonormal set. In particular, we note that the
digitisation process itself may lead to a mode set that does not
inherit the properties of the continuous mode functions from which
they are derived.  Therefore, considerable care must be exercised when
one performs decompositions using digitised versions of continous mode
functions that form orthonormal sets. Adopting the standard approach
without first investigating the properties of the digitised modes can
lead to unnecessarily inaccurate image decompositions.

\begin{table}
\begin{center}
\begin{tabular}{ccc}
\hline
Image & $\epsilon$ (standard) & $\epsilon$ (SVD) \\
\hline
1 & $2.923 \times 10^{-3}$ & $1.355 \times 10^{-3}$ \\
2 & $2.688 \times 10^{-3}$ & $1.938 \times 10^{-3}$ \\
3 & $1.519 \times 10^{-3}$ & $1.073 \times 10^{-3}$ \\
4 & $1.136 \times 10^{-3}$ & $9.483 \times 10^{-4}$ \\
5 & $2.235 \times 10^{-3}$ & $1.165 \times 10^{-3}$ \\
\hline
\end{tabular}
\caption{The residuals $\epsilon$ of the decompositions of the galaxy
  images shown in Fig.\ref{fig13} for the standard method and the
  SVD method.}
\label{tab1}
\end{center}
\end{table}

We illustrate our general approach by first applying it to the
decomposition of simple one-dimensional functions into Gauss-Hermite
(shapelet) modes. We show that, although the continuous shapelet mode
functions form an orthonormal set on the entire real line, the
corresponding digitised mode vectors can be linearly-dependent and
non-orthogonal, depending on the relative values of between the scale
parameter $\beta$, the pixel size and the size of the image under
analysis. We show that, for a wide range of these values, the method
developed here produces decompositions that are significantly superior
to those of the standard approach. In some cases, the decomposition
residual for our method is many orders of magnitude lower than that
for the standard technique.

We also illustrate our method by decomposing images of galaxies
extracted from the Hubble Deep Fields into two-dimensional
Gauss-Laguerre modes. We consider both the polar shapelet approach where
the modes are indexed using the energy and angular momentum quantum
numbers $n$ and $m$, and the $pm$-modes used in optics, which are
indexed by a radial number $p$ and azimuthal number $m$. 
In both cases, although the set of
continuous functions is orthonormal over the entire Euclidean plane,
the correspondinig digitised modes can be severely non-orthogonal, 
depending on the relative sizes of the mode
functions, the image size and the pixel scale. We show that the SVD method
proposed here again outperforms the standard approach, yielding
more faithful decompositions with lower residuals. We would therefore
advise that in future applications of the shapelet decomposition
method, the standard approach should be replaced with that presented
here. We also find that the $pm$-mode set provides consistently more
accurate decompositions than the polar shapelets.

The decompositions presented in this paper used the SVD to provide an
optimal decomposition for individual images. Frequently in astronomy,
an analysis of the statistics of a collection of images is used to
classify astronomical sources and properties. For such an application
it may be desirable to use the duals method we have presented. In this
case, the set of dual vectors can be calculated using a SVD, without
reference to any image. The expansion coefficients for each image can
then be very efficiently calculated by performing scalar products with
the duals. Additionally, at this stage, a priori knowledge can be used
(if desired) to constrain the duals to preferentially extract certain
features from any image.

We conclude that the ability to work straightforwardly with
linearly-dependent, non-orthogonal mode sets can be useful not just in
accommodating digitisation effects. In many applications, it may be
more efficient to work with modes that are known a priori to have such
properties. For example, in one or two dimensions, it can be prove
useful to decompose images simultaneously in shapelet mode sets
centred at a number of points in the image. The `multi-centre'
shapelet bases will be discussed in a forthcoming paper.  Finally we
draw the readers attention to our own previous work investigating
generalised bases or arbitrary completeness and orthogonality within
the field of THz optics (Berry et al. 2003; Withington et al. 2002),
including the description of second-order statistics associated with 
partially-coherent fields.

\subsection*{ACKNOWLEDGMENTS}
RHB acknowledges the support of a PPARC studentship, the Cavendish
Laboratory, the Cambridge Philosophical Society and Pembroke College,
Cambridge. The authors thank Dr. Phil Marshall for providing the
images of HDF galaxies.

\appendix

\section{$pm$-modes and polar shapelets}

\begin{figure}
\begin{center}
\includegraphics[width=0.95\linewidth]{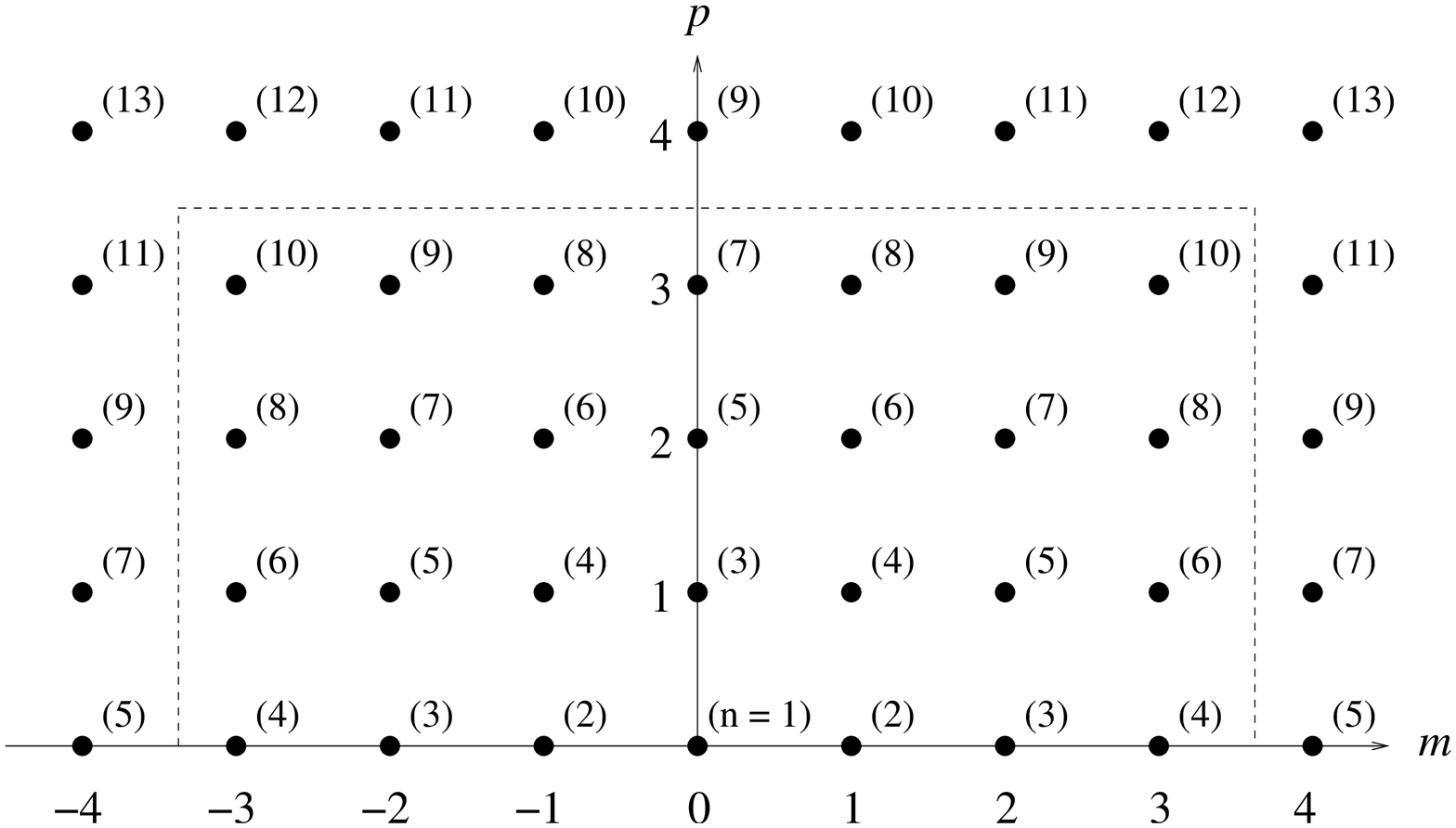}
\caption{The distributions of mode functions in the $pm$-plane. The
  numbers in brackets denote the value of the energy quantum number
  $n$ for each mode. The dashed line indicates the usual geometry
adopted in the $pm$-plane for truncating the mode set; in this case
$p_{\rm max} = |m|_{\rm max} = 3$.}
\label{figa1}
\end{center}
\end{figure}

The Schr\"odinger equation for the two-dimensional isotropic
harmonic oscillator may be written in plane polar coordinates as
\[
-\sspd{\psi}{r} - \frac{1}{r}\pd{\psi}{r}-\frac{1}{r^2}\sspd{\psi}{r}
+ r^2 \psi = 2E\psi,
\]
where we have set $\hbar=m=1$ and chosen our radial coordinate such
that the potential is given by $V(r)=\frac{1}{2}r^2$. Seeking a
separated solution of the form $\psi(r,\theta) = R(r)\Theta(\theta)$,
one quickly finds the set of solutions
\[
\psi_p^m(r,\theta) = A \exp(-r^2/2) (r^2)^{|m|/2} L_p^{|m|}(r^2) 
\exp(\mbox{i}m\theta),
\]
where  $L_p^{|m|}(x)$ is an associated Laguerre polynomial
and $A$ is a normalisation constant. 
In addition to being energy eigenstates, these solutions are clearly also
eigenstates of the angular momentum operator
$-\mbox{i}\partial/\partial\theta$ with eigenvalue $m$.
The angular momentum quantum number may take the values
$m = 0,\pm 1, \pm2, \ldots, \pm \infty$, whereas the
`radial' quantum number may take the values $p=0,1,2,\ldots,\infty$.
Fig.~\ref{figa1} shows the mode function distribution in the $pm$-plane.

Interestingly, the functions (\ref{nm-decomp}) are also the solutions of the 
paraxial wave equation in optics (see e.g. Goldsmith 1998), where they
are called the Gaussian beam $pm$-modes or simply $pm$-modes. These
modes are often used to describe the electric field distribution
in paraxial quasi-optical systems. In this context, it is customary
to truncate the mode set by imposing both a 
$p_{\rm  max}$-value and a $|m|_{\rm max}$ value. Moreover, it is
usual to choose $|m|_{\rm max} = p_{\rm max}$, so the truncated
set of modes form a rectangular in $pm$-space, as illustrated in
Fig.~\ref{figa1}. In this way, one obtains a total number of modes
given by 
\[
{\cal N}_1 = (p_{\rm max}+1)(2p_{\rm max}+1).
\]

It is straightforward to show that the energy of the mode $\psi_p^m$
is given by (reinstating $\hbar$ for the moment)
\[
E=(2p+|m|+1)\hbar\equiv (n+1) \hbar,
\]
where $n$ is the energy quantum number. 
Indeed, in the context of the two-dimensional harmonic
oscillator, it is more usual to label the modes in terms of $n$ and
$m$, rather than $p$ and $m$. In this way one arrives at the
polar shapelet modes
\[
\phi_n^m(r,\theta) = B \exp(-r^2/2) (r^2)^{|m|/2} 
L_{\frac{n-|m|}{2}}^{|m|}(r^2) 
\exp(\mbox{i}m\theta),
\]
where $B$ is a normalisation constant. The energy quantum number 
can take the values $n=0,1,2,\ldots,\infty$, whereas the azimuthal
(angular momentum) quantum number takes the values
$m = -n, -n+2, \ldots, n-2, n$; hence the energy level with
quantum number $n$ is
$(n+1)$-fold degenerate. Since we have performed just a simple
relabelling, the polar shapelets
(\ref{nm-decomp}) and the $pm$-modes have the same functional forms, and
are directly related by $\psi_p^m = \phi_{2p+|m|}^m$.
\begin{figure}
\begin{center}
\includegraphics[width=0.95\linewidth]{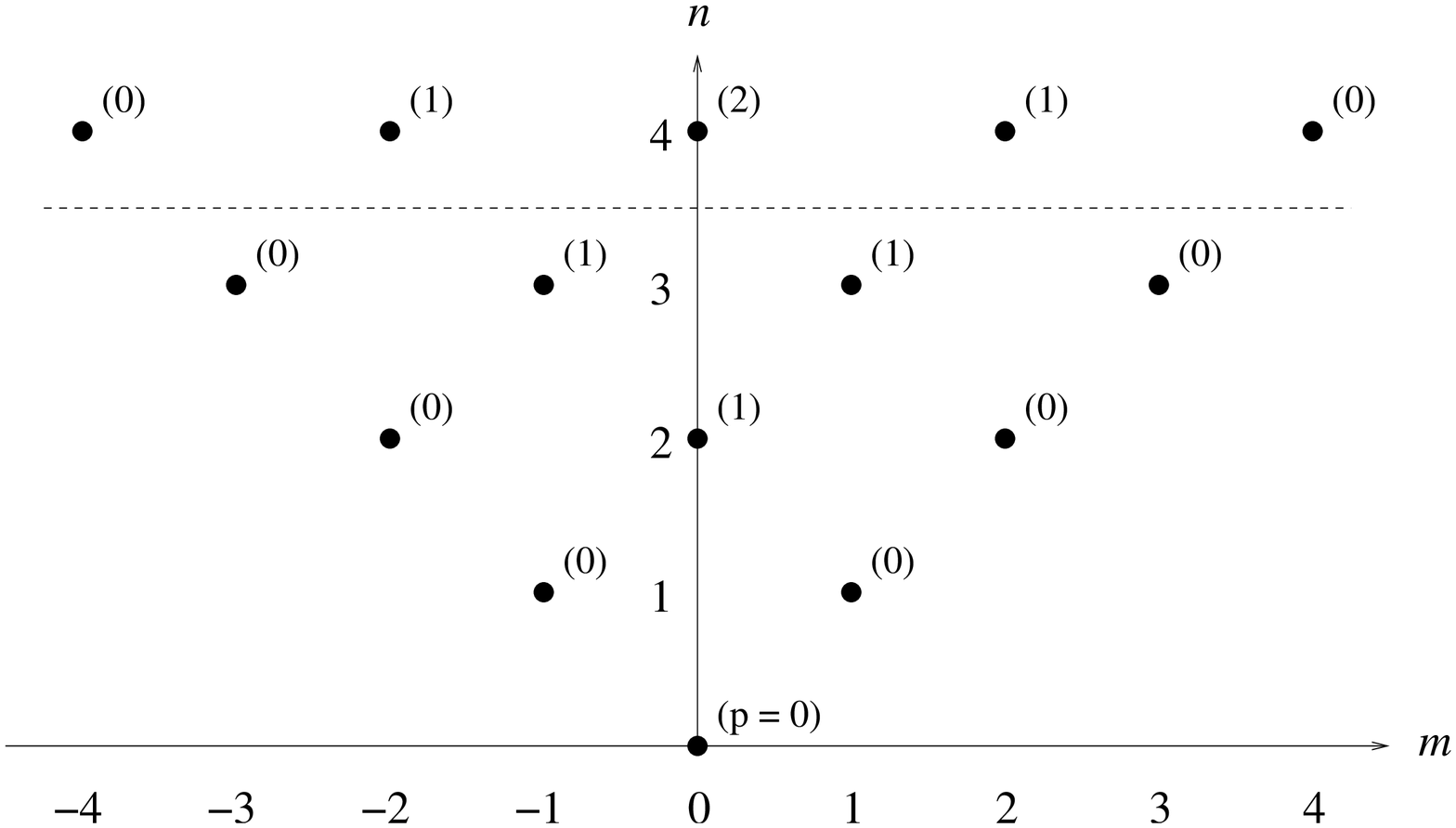}
\caption{The distributions of mode functions in the $nm$-plane. The
  numbers in brackets denote the value of the radial quantum number
  $p$ for each mode and indeed may be a more intuitive index than $n$ when investigating images due to $p$ being an intrinisic measure of radial extent and scale. The dashed line indicates the usual geometry
adopted in the $nm$-plane for truncating the mode set; in this case
$n_{\rm max} = 3$.}
\label{figa2}
\end{center}
\end{figure}
In Fig.~\ref{figa2} we show the distribution of modes in the
$nm$-plane. In Refregier (2003) and Massey et al. (2003),
to perform the decomposition of two-dimensional images, the mode
set is truncated by imposing an $n_{\rm max}$-value, as illustrated in
Fig.~\ref{figa2}. In this way, one obtains a total number of modes given by
\[
{\cal N}_2 = \tfrac{1}{2}(n_{\rm max}+1)(n_{\rm max}+2).
\]

\bsp 
\label{lastpage}
\end{document}